\documentstyle{article}

\begin{document}

\baselineskip=23pt

\begin{flushleft}
{\bf {\Huge A simple method of determining the Hubble constant}} 
\footnote{This essay received an ``honorable mention'' in the 1999 Essay Competition
of the Gravity Research Foundation.}
\\

{\bf Yi-Ping Qin$^{1,2,3,4}$ }

{\bf $^{1}$ Yunnan Observatory, Chinese Academy of Sciences, Kunming, Yunnan
650011, P. R. China; E-mail: ypqin@public.km.yn.cn }

{\bf $^{2}$ National Astronomical Observatories, Chinese Academy of Sciences 
}

{\bf $^{3}$ Chinese Academy of Science-Peking University joint Beijing
Astrophysical Center }

{\bf $^{4}$ Yunnan Astrophysics Center } \\
\end{flushleft}

\begin{center}
{\bf {\Large Abstract}}
\end{center}
\begin{quote}
Bidirectional relativistic proper motions of radio components of nearby 
extragalactic sources give a strong constraint on the determination of the 
Hubble constant $H_0$. Under the assumption that the real velocity of radio 
components of extragalactic sources is not less than that of Galactic sources, 
the value of $H_0$ can be estimated at a high level of accuracy. The assumption 
is reasonable due to the general belief that the activity in the core of 
galaxies must be more powerful than that of stars. This method is simple and 
with only one uncertainty --- the real velocity of components. This uncertainty 
is related to the value of the real velocity of componenets of Galactic sources 
and the latter is always well-determined (note that the determination is 
independent of $H_0$ and the distance of Galactic sources can be directly 
measured at a rather high level of accuracy). Hopefully the method will play 
an important role in future research to fix the value of $H_0$. With the data 
of the three sources available so far and the assumption that the real velocity 
of componenets of at least one of the sources is not less than a known velocity 
of componenets of a Galactic source, the constant is estimated to be within 
$27.08kms^{-1}Mpc^{-1} < H_0 \leq 53.15kms^{-1}Mpc^{-1}$ with this method.
\\
\end{quote}

\vspace{2mm}

The real value of the Hubble constant has been a hot topic for a long time. 
In the past few years, some advances have been achieved (for more details, 
see Trimble and Aschwanden 1999, Trimble and McFadden 1998). An exciting 
result of observation by HST led to $H_0 = 80 \pm 17 kms^{-1}Mpc^{-1}$ 
(Freedman et al. 1994). However, with the same data, some people got a 
much smaller value, e. g., $H_0 = 40 kms^{-1}Mpc^{-1}$ (Sandage et 
al. 1994). Among the many methods, that taking the peak brightness of Type
Ia supernovae as a distance indicator is generally used. It gave a small
value of the constant, $H_0 = 61 \pm 10 kms^{-1}Mpc^{-1}$, by Brach (1992).

Bidirectional relativistic motions of extragalactic radio sources can be
used to estimate the constant (Marscher and Broderick 1982). Recently there 
were some sound works using this method published. The most successful one
was done by Taylor et al. (1997). However, the method they used is 
somewhat complicated. And it does not tell how the uncertainties of the 
measurements used affect the estimation of $H_0$, and what one should 
do when the data of several sources are available. 

In the following, we illustrate a simple method of determining the Hubble
constant using bidirectional relativistic proper motions of extragalactic
radio sources.

The apparent transverse velocities of components of a source along an 
axis at an angle $\theta $ to the line of sight at a velocity $\beta c$ can 
be expressed as (Rees 1966, 1967)
\begin{equation}
(\beta _{app})_a\equiv \frac{\mu _aD_L}{c(1+z)}=\frac{\beta \sin \theta }{%
1 - \beta \cos \theta }, 
\end{equation}
\begin{equation}
(\beta _{app})_r\equiv \frac{\mu _rD_L}{c(1+z)}=\frac{\beta \sin \theta }{%
1 + \beta \cos \theta }, 
\end{equation}
where $a$ and $r$ stand for the motions of the approaching and receding 
components, with $\mu _a$ and $\mu _r$ being the corresponding
proper motions, respectively. 

These two equations give
\begin{equation}
\frac{D_L}{c(1+z)}=\frac 1{2\mu _a\mu _r}\sqrt{\beta ^2(\mu _a+\mu
_r)^2-(\mu _a-\mu _r)^2}. 
\end{equation}
For small redshift sources, the following is maintained for all kinds of 
the universe within the framework of the FRW cosmology, 
\begin{equation}
\frac{D_L}{1+z}\simeq \frac{cz}{H_0}, 
\end{equation}
where $H_0$ is the Hubble constant of the universe. 

From Equation (3) one has 
\begin{equation}
H_0\simeq \frac{2\mu _a\mu _rz}{\sqrt{\beta ^2(\mu _a+\mu _r)^2-(\mu _a-\mu
_r)^2}}. 
\end{equation}

It shows that, for a nearby extragalactic source with measured
values of $z$, $\mu _a$ and $\mu _r$, when $\beta $ is known, the Hubble
constant would be well determined. This relation provides a very strong
constrain on the determination of $H_0$. 

From Equation (5), the law of $\beta <1$ leads to 
\begin{equation}
H_0>z\sqrt{\mu _a\mu _r} 
\end{equation}
for any sources. Therefore, among many values of the lower limit of $H_0$,
calculated from various sources, the largest one would be the closest value
to the limit. If $\mu _a$ and $\mu _r$ are presented in the form $\mu \pm
\bigtriangleup \mu $ for a source, then the lower limit of $H_0$ determined
by the source should be 
\begin{equation}
H_{0,\min }=z\sqrt{(\mu _a-\bigtriangleup \mu _a)(\mu _r-\bigtriangleup \mu
_r)}, 
\end{equation}
with $H_{0,\min }$ satisfying 
\begin{equation}
H_{0,\min }<H_0. 
\end{equation}
In this way, the largest value of $H_{0,\min }$ among those calculated from
various sources should be taken as the best estimation of the lower limit of 
$H_0$.

Let 
\begin{equation}
\alpha \equiv 1-\beta . 
\end{equation}
Since $0\leq \beta <1$, then $0<\alpha \leq 1$. For a small value of $\alpha 
$, Equation (5) gives 
\begin{equation}
H_0\simeq z\sqrt{\mu _a\mu _r}[1+\frac{(\mu _a+\mu _r)^2}{4\mu _a\mu _r}%
\alpha ]. 
\end{equation}
Considering the case where $\alpha $ is known to the extent of $\alpha \leq
\alpha _{\max }$ for a source, the upper limit of the Hubble constant
would be determined by the source in the way 
\begin{equation}
H_0\leq H_{0,\max }, 
\end{equation}
where 
\begin{equation}
H_{0,\max }=z\sqrt{(\mu _a+\bigtriangleup \mu _a)(\mu _r+\bigtriangleup \mu
_r)}[1+\frac{(\mu _a+\bigtriangleup \mu _a+\mu _r+\bigtriangleup \mu _r)^2}{%
4(\mu _a+\bigtriangleup \mu _a)(\mu _r+\bigtriangleup \mu _r)}\alpha _{\max
}]. 
\end{equation}

In determination of the range of $H_0$, there is a reasonable demand that
the ranges of $H_0$ estimated from different sources should be overlapped.
This demand is consistent with the above principle of choosing the lower
limit of $H_0$. When determining the upper limit of $H_0$ from (12), the
requirement can be realized by adopting various values of $\alpha _{\max }$
for different sources.

If among these sources, there is at least one source satisfying $\alpha \leq
\alpha _{\max }$ for a given $\alpha _{\max }$, the largest value of $%
H_{0,\max }$ calculated with this value of $\alpha _{\max }$ for all the
sources should be taken as the best estimation of 
the upper limit of $H_0$. In this way, while the
given value of $\alpha _{\max }$ is assigned to the source of the largest
value of $H_{0,\max }$, some larger values of $\alpha _{\max }$ should
be assigned to other sources, so that the estimated ranges of $H_0$ may be
overlapped.

Since the value of $\alpha _{\max }$ for Galactic sources can be calculated
at a rather high level of accuracy, that for extragalactic sources can then
be well settled by assuming that it would not be less than that for Galactic
sources. This assumption is reasonable due to the general belief that the
activity in the core of galaxies must be more powerful than that of stars.

Recently, several Galactic sources were found to have
bidirectional relativistic proper motions of radio components. Among them,
the largest and well calculated value of $\beta $ is $0.92\pm 0.02$ for GRO
J1655-40 (Hjellming and Rupen 1995). 
This corresponds to $\beta _{\min }=0.9$ and $\alpha _{\max
}=0.1$ for the source. Therefore, at present, it is reasonable to take $%
\alpha _{\max }=0.1$ for extragalactic sources according to the assumption.

Till now, there are only a few extragalactic sources with measured values of 
$\mu _a$ and $\mu _r$ found in literature. Excluding those with high
redshifts or uncertain values of proper motions, there are only three
sources suitable for our study. They are: 1146+596 (NGC 3894), $%
z=0.01085$, $\mu _a=0.26\pm 0.05masyr^{-1}$ and $\mu _r=0.19\pm 
0.05masyr^{-1}$ (Taylor et al. 1998); 0316+413 (3C 84), $z=0.0172$, 
$\mu _a=0.58\pm 0.12masyr^{-1}$ and $\mu _r\leq 0.28masyr^{-1}$ 
(Marr et al. 1989, Vermeulen et al. 1994); 1946+708, $z=0.101$, 
$\mu _a=0.117\pm 0.020masyr^{-1}$ and $\mu _r=0.053\pm 0.020masyr^{-1}$ 
(Taylor and Vermeulen 1997).

For the lower limit of $H_0$, the first and the third sources give $%
H_{0,\min }=8.82kms^{-1}Mpc^{-1}$ and $27.08kms^{-1}Mpc^{-1}$, respectively
from (7), while the second source gives no lower limit of $H_0$. According 
to the above principle of choosing $H_{0,\min }$, the best estimation of the 
lower limit of $H_0$ from these data should be 
$H_{0,\min }=27.08kms^{-1}Mpc^{-1}$.

For the given value of $\alpha _{\max }=0.1$, the three sources give $%
H_{0,\max }=15.45kms^{-1}Mpc^{-1}$, $40.51kms^{-1}Mpc^{-1}$, and $%
53.15kms^{-1}Mpc^{-1}$, respectively from (12). Assuming that there
is at least one source satisfying $\alpha \leq \alpha _{\max }$ for 
$\alpha _{\max }=0.1$, then according to the above requirement we
choose $H_{0,\max }=53.15kms^{-1}Mpc^{-1}$ as the best estimation of the
upper limit of $H_0$.

Therefore, we obtain the range of $27.08kms^{-1}Mpc^{-1}<H_0\leq
53.15kms^{-1}Mpc^{-1}$ for the Hubble constant from the data of the three
sources.

In practice, the observable bidirectional relativistic proper motions of
radio components of a source are always those moving almost perpendicular to
the line of sight, and then the values of their $\mu _a$ and $\mu _r$ are
close (see, e.g., Taylor et al. 1998). Therefore $\frac{%
(\mu _a+\bigtriangleup \mu _a+\mu _r+\bigtriangleup \mu _r)^2}{4(\mu
_a+\bigtriangleup \mu _a)(\mu _r+\bigtriangleup \mu _r)}\simeq 1$ and $\frac{%
(\mu _a+\bigtriangleup \mu _a+\mu _r+\bigtriangleup \mu _r)^2}{4(\mu
_a+\bigtriangleup \mu _a)(\mu _r+\bigtriangleup \mu _r)}\alpha _{\max
}\simeq 0.1$ for $\alpha _{\max }=0.1$. It shows that taking $\alpha
_{\max }=0.1$ will only produce about $10\%$ uncertainty for $H_{0,\max }$.
When more such sources have been observed, the expected value of $\alpha
_{\max }$ and the corresponding uncertainty will be smaller. 

This method depends on only one assumption and it concerns only one
uncertainty --- the real velocity of components. This uncertainty concerns
the value of the real velocity of components of Galactic sources and the
latter is always well determined (note that the determination is independent
of $H_0$ and the distance of Galactic sources can be directly measured at a
rather high level of accuracy). The assumption is weak due to the general
belief that the activity in the core of galaxies must be more powerful than
that of stars. Also, the method is simple. To determine $H_0$ at a high
level, one only needs to measure $\mu _a$ and $\mu _r$ at a high level of
accuracy and finds a sufficient number of such sources (say, 10 or more). 
The method is then hopeful to play an important role for finally fixing the 
value of the Hubble constant in future researches.

\vspace{20mm}

{\bf ACKNOWLEDGEMENTS}

It is a pleasure to thank G. B. Taylor, R. C.
Vermeulen and Y. H. Zhang for kindly supplying necessary materials. This
work was supported by the United Laboratory of Optical Astronomy, CAS, the
Natural Science Foundation of China, and the Natural Science Foundation of
Yunnan.

\newpage

\begin{verse}
{\bf REFERENCES}

Brach, D. 1992, {\it Astrophys. J.}, {\bf 392}, 35\\

Freedman, W. L. et al. 1994, {\it Nature}, {\bf 371}, 757\\

Hjellming, R. M., and Rupen, M. P. 1995, {\it Nature.}, {\bf 375}, 464\\

Marr, J. M., Backer, D. C., and Wright, M. C. H. et al. 1989, {\it Astrophys. 
J.}, {\bf 337}, 671\\

Marscher, A. P., and Broderick, J. J. 1982, In {\it Extragalactic Radio 
Sources}, D. S. Heeschen and C. M. Wade, eds. (Dordrecht: Reidel), p. 359\\

Rees, M. J. 1966, {\it Nature}, {\bf 211}, 468\\

Rees, M. J. 1967, {\it Mon. Not. R. Astron. Soc.}, {\bf 289}, 945\\

Sandage, A. et al. 1994, {\it Astrophys. J.}, {\bf 423}, L13\\

Taylor, G. B., and Vermeulen, R. C. 1997, {\it Astrophys. J.}, {\bf 485}, 
L9\\

Taylor, G. B., Worbel, J. M., and Vermeulen, R. C. 1998, {\it Astrophys. 
J.}, {\bf 498}, 619\\

Trimble, V., and Aschwanden, M. 1999, {\it Pub. Astron. Soc. Pacific}, 
{\bf 111}, 385\\

Trimble, V., and Mcfadden, L.-A. 1998, {\it Pub. Astron. Soc. Pacific}, 
{\bf 110}, 223\\

Vermeulen, R. C., Readhead, A. C. S., and Backer, D. C. 1994, {\it Astrophys. 
J.}, {\bf 430}, L41\\

\end{verse}

\end{document}